\documentstyle[12pt,epsfig]{article}

\advance\textheight by \topskip
\begin{document}
\tolerance=100000
\thispagestyle{empty}
\setcounter{page}{1}

\newcommand{\be}{\begin{equation}}
\newcommand{\ee}{\end{equation}}
\newcommand{\br}{\begin{eqnarray}}
\newcommand{\er}{\end{eqnarray}}
\newcommand{\ba}{\begin{array}}
\newcommand{\ea}{\end{array}}
\newcommand{\bi}{\begin{itemize}}
\newcommand{\ei}{\end{itemize}}
\newcommand{\bn}{\begin{enumerate}}
\newcommand{\en}{\end{enumerate}}
\newcommand{\bc}{\begin{center}}
\newcommand{\ec}{\end{center}}
\newcommand{\ul}{\underline}
\newcommand{\ol}{\overline}
\newcommand{\ar}{\rightarrow}
\newcommand{\sm}{${\cal {SM}}$}
\newcommand{\mssm}{${\cal {MSSM}}$}
\newcommand{\susy}{{{SUSY}}}
\def\epem{\ifmmode{e^+ e^-} \else{$e^+ e^-$} \fi}
\newcommand{\Dir}{\kern -6.4pt\Big{/}}
\newcommand{\Dirin}{\kern -10.4pt\Big{/}\kern 4.4pt}
\newcommand{\DDir}{\kern -7.6pt\Big{/}}
\newcommand{\DGir}{\kern -6.0pt\Big{/}}
\newcommand{\eett}{$e^+e^-\rightarrow t\bar t $}
\newcommand{\eeZphi}{$e^+e^-\rightarrow Z\phi $}
\newcommand{\eeZH}{$e^+e^-\rightarrow ZH $}
\newcommand{\eeAH}{$e^+e^-\rightarrow AH $}
\newcommand{\eehww}{$e^+e^-\rightarrow hW^+W^-$}
\newcommand{\bbww}{$b\bar b W^+W^-$}
\newcommand{\eebbww}{$e^+e^-\rightarrow b\bar b W^+W^-$}
\newcommand{\ttbbww}{$t\bar t\rightarrow b\bar b W^+W^-$}
\newcommand{\eettbbww}{$e^+e^-\rightarrow t\bar t\rightarrow b\bar b W^+W^-$}
\newcommand{\bbb}{$ b\bar b$}
\newcommand{\ttb}{$ t\bar t$}
\def\Ord{\buildrel{\scriptscriptstyle <}\over{\scriptscriptstyle\sim}}
%
\def\OOrd{\buildrel{\scriptscriptstyle >}\over{\scriptscriptstyle\sim}}
\def\Ord{\buildrel{\scriptscriptstyle <}\over{\scriptscriptstyle\sim}}
\def\OOrd{\buildrel{\scriptscriptstyle >}\over{\scriptscriptstyle\sim}}
\def\pl #1 #2 #3 {{\it Phys.~Lett.} {\bf#1} (#2) #3}
\def\np #1 #2 #3 {{\it Nucl.~Phys.} {\bf#1} (#2) #3}
\def\zp #1 #2 #3 {{\it Z.~Phys.} {\bf#1} (#2) #3}
\def\pr #1 #2 #3 {{\it Phys.~Rev.} {\bf#1} (#2) #3}
\def\prep #1 #2 #3 {{\it Phys.~Rep.} {\bf#1} (#2) #3}
\def\prl #1 #2 #3 {{\it Phys.~Rev.~Lett.} {\bf#1} (#2) #3}
\def\mpl #1 #2 #3 {{\it Mod.~Phys.~Lett.} {\bf#1} (#2) #3}
\def\rmp #1 #2 #3 {{\it Rev. Mod. Phys.} {\bf#1} (#2) #3}
\def\sjnp #1 #2 #3 {{\it Sov. J. Nucl. Phys.} {\bf#1} (#2) #3}
\def\cpc #1 #2 #3 {{\it Comp. Phys. Comm.} {\bf#1} (#2) #3}
\def\xx #1 #2 #3 {{\bf#1}, (#2) #3}
\def\preprint{{\it preprint}}

\begin{flushleft}
{RAL-TR-99-071}
\end{flushleft}
\begin{flushright}
\vskip-1.0truecm
{LC-TH-2000-004}\\
{November 1999}
\end{flushright}
\vskip0.1cm\noindent
\begin{center}
{\Large {\bf Higgs radiation off top-antitop pairs at
 future Linear Colliders: a background study\footnote{Talk 
given at the 2nd ECFA/DESY Study on Physics 
and Detectors for a Linear Electron-Positron Collider, Oxford,
UK, 20-23 March 1999.}}}\\[0.5cm]
{\large S.~Moretti}\\[0.05 cm]
{\it Rutherford Appleton Laboratory, Chilton, Didcot, Oxon OX11 0QX, UK.}
\end{center}
\begin{abstract}
{\small
\vskip-0.20cm
\noindent
The process $e^+e^-\ar H t\bar t$ can be exploited 
at future Linear Colliders (LCs) \cite{ee500}--\cite{newee500}
to measure the Higgs-top Yukawa coupling.
In this note, we estimate the size of the irreducible backgrounds
in the channel $H t\bar t$ $\to$ $b\bar b b\bar b W^+W^-$ $\ar$ 
$b\bar b b\bar b \ell^\pm\nu_\ell$ $q\bar q'$, for the case of a 
Standard Model (SM) Higgs boson with mass between 100 and 140 GeV.
}
\end{abstract}
{\large\bf 1. Interplay of signal and backgrounds}
\vskip0.15cm\noindent
Higgs-strahlung off top-antitop pairs \cite{eetth} 
is the most promising channel to
measure the Yukawa coupling of the top quark to a Higgs boson \cite{top}.
In the SM, the production rate of this final state at future LCs is sizable, 
if the energy of the latter is of the order $\sqrt s\OOrd 2m_t+M_H$.
Furthermore, the one-loop QCD corrections to $e^+e^-\ar H t\bar t$ 
are under control \cite{nlo}.
Thus, in the context of simulation studies, the next thing to be assessed is
the viability of the signal above the backgrounds.

The current experimental limit from LEP2 on $M_H$ is about 95 GeV 
\cite{limitH}. In addition, electroweak  fits to precision data
prefer a rather light Higgs boson, say, below 150 GeV within the SM 
\cite{blois}. Thus, a LC with centre-of-mass (CM)
energy $\sqrt s=500$ GeV (the default of our numerical studies)
represents an ideal laboratory to study
$Ht\bar t$ final states (as $m_t=175$ GeV).
In the mass range 90 GeV $\Ord  M_H\Ord $ 150 GeV, the
dominant Higgs decay mode is $H\ar b\bar b$, this being overtaken by the 
off-shell decay into two $W^\pm$'s, i.e., $H\ar W^{+*}W^{-*}$, only for 
$M_H\OOrd 140$ GeV \cite{WJSZK}. Besides, the 
experimentally preferred channel in searching
for $t\bar t$ events is  the semi-leptonic channel, i.e.,
$t\bar t \ar b\bar b W^+W^-\ar b\bar b \ell^\pm\nu_\ell$ $q\bar q'$, where
$\ell$ and $\nu$ represent a lepton at high transverse momentum
(used for triggering purposes) and its companion neutrino
and $q\bar q'$ refers to all possible combinations of light quarks.
These are the signatures of Higgs and top-antitop pairs that 
we will  concentrate on here. A study of `reducible' backgrounds
in the above decay channels was preliminarly
presented in \cite{merino}: there, they
were found to be under control after appropriate
selection cuts were implemented.
In this note we consider the effects of the dominant 
`irreducible' backgrounds, i.e., those of the type \cite{first}--\cite{eettg}: 
\begin{equation}\label{htt_hbbww}
e^+e^-\ar H t\bar t\ar H b\bar b W^+W^-\ar b\bar b b\bar b \ell^\pm
\nu_\ell q\bar q',
\end{equation}
\begin{equation}\label{ztt_zbbww}
e^+e^-\ar Z t\bar t\ar Z b\bar b W^+W^-\ar b\bar b b\bar b \ell^\pm
\nu_\ell q\bar q',
\end{equation}
\begin{equation}\label{gtt_gbbww}
e^+e^-\ar g t\bar t\ar g b\bar b W^+W^-\ar b\bar b b\bar b \ell^\pm
\nu_\ell q\bar q',
\end{equation}
whose corresponding Feynman graphs can be found in
Figs.~1--3 of Ref.~\cite{io} (the latter to be further 
supplemented with the $H,Z$ and $g$ decay currents) as well as those
of all other subleading reactions at the same perturbative orders,
\begin{equation}\label{hbbww}
e^+e^-\ar b\bar b W^+W^- \ar b\bar b b\bar b \ell^\pm
\nu_\ell q\bar q',
\end{equation}
\begin{equation}\label{zbbww}
e^+e^-\ar Z b\bar b W^+W^- \ar b\bar b b\bar b \ell^\pm
\nu_\ell q\bar q',
\end{equation}
\begin{equation}\label{gbbww}
e^+e^-\ar g b\bar b W^+W^- \ar b\bar b b\bar b \ell^\pm
\nu_\ell q\bar q'.
\end{equation}
In total, one
has to consider at  tree-level 350 Feynman graphs for process (\ref{hbbww}),
546 for reaction (\ref{zbbww}) and 152 for case ({\ref{gbbww}), each
number including the graphs associated to 
(\ref{htt_hbbww})--(\ref{gtt_gbbww}). 
Their computation has been accomplished by using helicity amplitude techniques
to evaluate the complete 
$2\ar 8$ body reactions, without any factorisation of production
and decay processes \cite{io}.
\vskip0.25cm\noindent
{\large\bf 2. Cross sections}
\vskip0.15cm\noindent
If one does assume the mentioned Higgs and top-antitop decay modes, then
signal  events can be searched for in data samples 
made up by four $b$ quark jets, two light quark jets,
a lepton and a neutrino. In other terms, 
a `$4b + 2~{\mathrm{jets}}~+ \ell^\pm + E_{\mathrm{miss}}$' signal,
 with $\ell=e,\mu,\tau$\footnote{We include
$\tau$'s to enhance the signal rate, assuming that they
are distinguishable from quark jets.}.
Fig.~\ref{fig:cross} presents the production cross sections
 for the leading subprocesses (\ref{htt_hbbww})--(\ref{gtt_gbbww})
as well as those for the complete ones (\ref{hbbww})--(\ref{gbbww}).
\begin{figure}[!t]
~\hskip3.0cm\epsfig{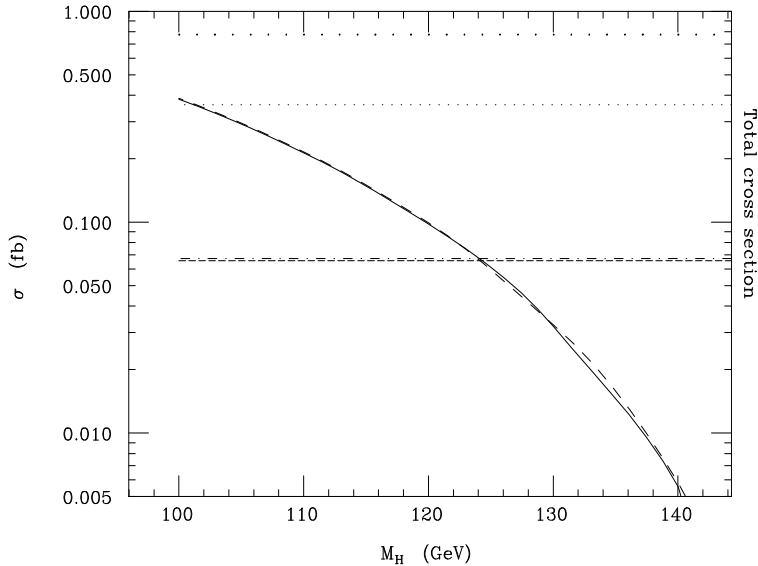}
\caption{\small Cross sections for the following processes (see the text): 
(\ref{htt_hbbww}) (solid), 
(\ref{ztt_zbbww}) (dashed), 
(\ref{gtt_gbbww}) (fine-dotted), 
(\ref{hbbww}) (long-dashed), 
(\ref{zbbww}) (dot-dashed) and 
(\ref{gbbww}) (dotted). 
}
\label{fig:cross}
\end{figure}
\noindent
{}From there, two aspects emerge. On the one hand, the QCD processes
are always dominant, whereas the interplay between the other two
depends upon the Higgs mass value. On the other hand, 
the bulk of the cross sections of processes
(\ref{hbbww})--(\ref{gbbww}) comes  from 
(\ref{htt_hbbww})--(\ref{gtt_gbbww}) with the only exception 
of the  QCD case.  The increase of the 
QCD rates seen in the figure is mainly 
due to the large amount of gluon radiation off $b$ (anti)quarks produced 
in `radiative' $t\bar t$ decays \cite{eettg}: i.e., via 
$t\bar t\ar b\bar b W^+W^-\ar g b\bar b
W^+W^-$, with the gluon eventually
yielding $b\bar b$ pairs: see Fig.~\ref{fig:talk_gb}.
\begin{figure}[!h]
~\hskip3.0cm\epsfig{file=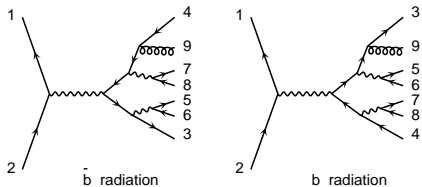,height=16cm,angle=0}
\vskip-13.5cm
\caption{\small The diagrams responsible for radiative top-antitop decays
in process (\ref{gbbww}).}
\label{fig:talk_gb}
\end{figure}
However, it is instructive to further explore the background effects in the
case of $Hb\bar b W^+W^-$ events. This is done in Fig.~\ref{fig:split_tth},
where we have distinguished between subprocess (\ref{htt_hbbww})
and four sizable components of (\ref{hbbww}), i.e., the cases proceeding
via intermediate states of the following forms: 
(\ref{hbbww}a) $H t\bar b W^- +~{\mathrm{c.c.}}$,
(\ref{hbbww}b) $H Z W^+W^-$,
(\ref{hbbww}c) $H \gamma W^+W^-$,
(\ref{hbbww}d) $H H W^+W^-$,
(\ref{hbbww}e) all remaining interferences. A notable aspect 
in Fig.~\ref{fig:split_tth} is the taking
over of the `single-top' production rates (\ref{hbbww}a) respect to the 
`double-top' ones (\ref{htt_hbbww}) for large Higgs masses, because of the 
strong phase space suppression on the latter when $M_H$ approaches 
$\sqrt s-2m_t$. Besides, (\ref{hbbww}a)  events carry the same Yukawa 
dependence as (\ref{htt_hbbww}), see Fig.~\ref{fig:bkgd_tth},
thus they should rather be regarded as an additional contribution to
the top-Higgs Yukawa signal, these two mechanisms
being somehow complementary with respect to the $M_H$ dependence.  

\begin{figure}[!t]
~\hskip3.0cm\epsfig{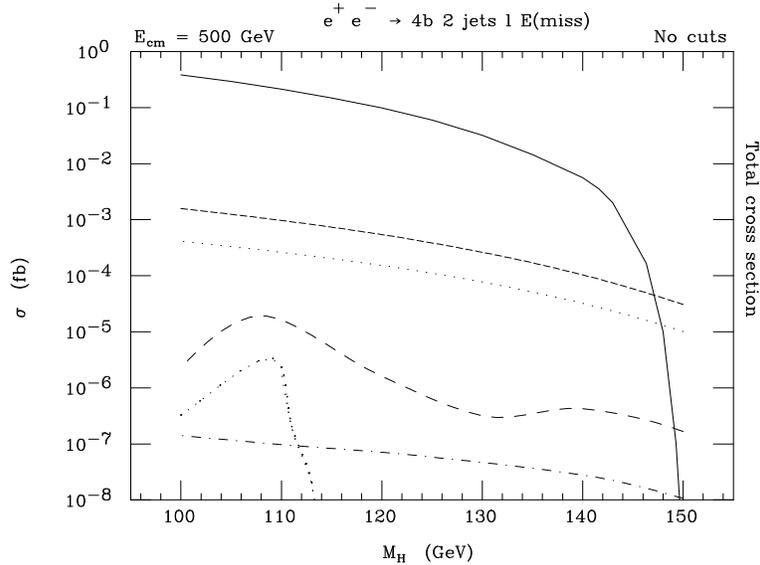}
\caption{\small Cross sections for the (\ref{htt_hbbww}) subprocess (solid)
and the following other components of the full reaction
(\ref{hbbww}) (see the text): 
(\ref{hbbww}a) (dashed), 
(\ref{hbbww}b) (dotted),
(\ref{hbbww}c) (long-dashed), 
(\ref{hbbww}d) (dot-dashed) and 
(\ref{hbbww}e) (fine-dotted). 
}
\label{fig:split_tth}
\end{figure}
\begin{figure}[!h]
~\hskip2.0cm\epsfig{file=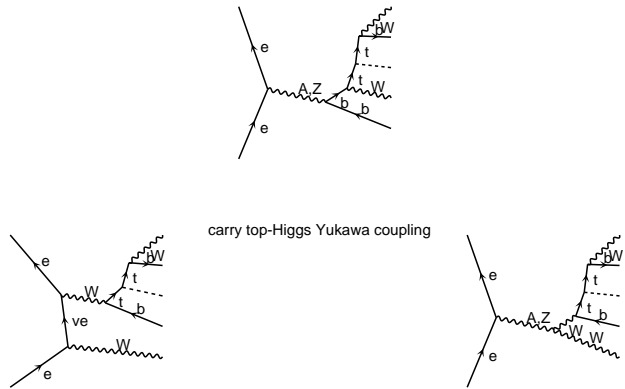,height=16cm,angle=0}
\vskip-10.5cm
\caption{\small The diagrams responsible for single-top events
in process (\ref{hbbww}).}
\label{fig:bkgd_tth}
\end{figure}

As one of the $b\bar b$ pairs in the final
state would naturally resonate at $M_H$, at $M_Z$ or logarithmically
increase at low mass, for processes
(\ref{htt_hbbww}), (\ref{ztt_zbbww}) and (\ref{gtt_gbbww}), respectively,
we investigate the di-jet mass spectra that can be reconstructed from
the four $b$ quarks in
the `$4b + 2~{\mathrm{jets}}~+ \ell^\pm + E_{\mathrm{miss}}$' signature.
Since we do not assume any jet-charge determination of 
the  $b$ jets and consider  negligible the mis-tagging 
of light-quark jets as heavy ones, six such combinations can be built up.
We distinguish among these by ordering the four $b$ jets in energy
(i.e., $E_1>E_2>E_3>E_4$),  in such a way that 
the $2b$ invariant mass $m_{ij}$ refers to the $ij$ pair (with
$i<j=2,3,4$) in which the $i$-th and $j$-th most energetic particles enter.
In Fig.~\ref{fig:mbb-Eb} (top), one can appreciate the
`resonant' shapes around $M_H$ and $M_Z$ in all $ij$ cases.
As for the `divergence' in the $g\ar b\bar b$ splitting of the QCD
process, this can easily be spotted in the case $ij=34$. In the end, the $2b$
mass spectra look rather promising as a mean of reducing
 both backgrounds (\ref{ztt_zbbww})--(\ref{gtt_gbbww}).
By requiring, e.g., $m_{34}>50$ GeV, one would vigorously reduce
the latter; similarly, by imposing, e.g., $|m_{14}-M_Z|>15$ GeV one would
reject the former considerably.

\begin{figure}[!t]
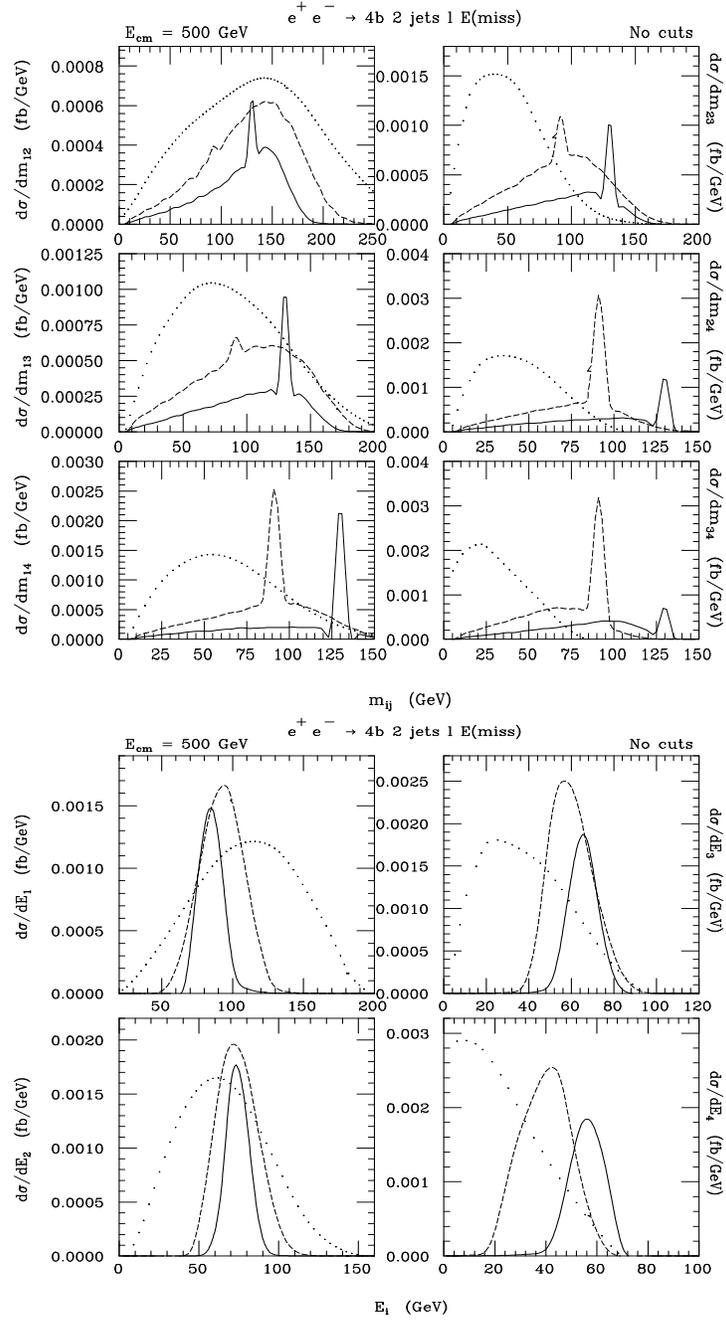

\begin{center}
~\epsfig{file=mbb.ps,width=9.5cm,height=9.5cm,angle=0}
\vskip0.0cm
~\epsfig{file=Eb.ps,width=9.5cm,height=8.0cm,angle=0}
\caption{\small Differential distributions in invariant mass
(top) and energy (bottom) of the
energy-ordered $b$ jets for the 
following processes (see the text):
(\ref{htt_hbbww}) (solid), 
(\ref{ztt_zbbww}) (dashed) and 
(\ref{gtt_gbbww}) (dotted). 
(Note that the rates of reaction (\ref{gtt_gbbww}) 
have been divided by three for readability.) 
}
\label{fig:mbb-Eb}
\end{center}
\end{figure}

An alternative way of looking at the same phenomenology is by studying the
energy spectra of the four $b$ quarks. In fact, the larger value of
$M_H$, as compared to $M_Z$, should boost the $b$ quarks generated by
the Higgs boson towards energies higher than those
achieved in the $Z$ decays. Conversely, the energy of the $b$ quarks
emerging from the two remaining unstable particles, top and antitop
quarks, should be softer in the first case. Following similar arguments,
one should expect the hardest(softest) $b$ (anti)quark from gluon events to
actually be the hardest(softest) of all cases
(\ref{htt_hbbww})--(\ref{gtt_gbbww}). Recalling that the two
most energetic $b$'s seldom come from a $H$, $Z$ or $g$ splitting, 
the above kinematic features are clearly 
recognisable in Fig.~\ref{fig:mbb-Eb} (bottom). Therefore, the energy spectra
too are useful in disentangling Higgs events.
If one imposes, e.g., 
$E_1<100$ GeV and $E_4>50$ GeV, both $Z$ and gluon events
can be strongly depleted, at a rather low cost for the signal.

\vskip0.25cm\noindent
{\large\bf 3. Conclusions}
\vskip0.15cm\noindent
Taking into account our results, we believe that
the study of the Higgs-top Yukawa coupling
at future $e^+e^-$ accelerators, running at
 500 GeV or higher, can be pursued by means of the 
Higgs-strahlung process $e^+e^-\ar Ht\bar t$. 
In particular, the irreducible backgrounds 
can be controlled in the decay channels 
$t\bar t\ar b\bar b W^+W^-\ar b\bar b \ell^\pm\nu_\ell q\bar q'$ and 
 $H\ar b\bar b$, for $M_H\Ord 140$ GeV.
Those presented here are predictions obtained 
at parton level only, though with an advanced perturbative
treatment based on $2\to8$ body processes.
They should be complemented with others at a more phenomenological
level, also including fragmentation/hadronisation and detector
effects. Indeed, several progress has recently been made in this respect
\cite{spainandusa}, which confirmed the feasibilty of this kind of analyses.
\vskip0.10cm
\noindent
\underbar{\sl Acknowledgements~} 
We thank the conveners
of the top working group for the
stimulating environment they have been able to create during the
workshops.
\end{document}